# InterPSS: A New Generation Power System Simulation Engine

Mike Zhou, *Member, IEEE*
State Grid EPRI
Beijing China
mike.zhou@interpss.org

Qiuhua Huang, Member, IEEE
Pacific Northwest National Laboratory
Richland, WA, USA
qiuhua.huang@pnnl.gov

*Abstract*—This paper presents the design of InterPSS simulation engine, including its object model, open software architecture, and software development process. Several advanced applications, including an integrated transmission and distribution co-simulation, an electromagnetic transient and phasor domain hybrid simulation, and InterPSS integration with a market simulator, have been developed by either extending InterPSS simulation engine or integrating it with other programs and/or platforms. These advanced applications show that the open architecture combined with the comprehensive modeling and simulation capabilities make InterPSS a very attractive option for the research and the future new power system simulation application development.

*Index Terms*--Hybrid simulation, Integrated transmission and distribution system co-simulation, Model-Driven Development, Power System Modeling, Power System Simulation.

## I. INTRODUCTION

### A. Motivation

The landscape of the IT technology has been fundamentally changed over the last 40 years. The computing environment available to the power system simulation has been changed from one workstation with one CPU to the computer cluster with multiple computer nodes, each with multiple CPU/Core. The software system architecture and the associated development process of power system simulation software should be also updated accordingly to catch-up the IT technology advances. Core simulation engines of the main stream commercial power system simulation software, such as PSS/E, BPA, and PSASP (developed by China Electric Power Research Institute), were mostly developed in the last century, mainly using the Fortran programming language and the procedure programming approach. Each application/program is mainly focusing on a specific problem or domain, e.g., AC Loadflow, Short Circuit analysis, transmission system or distribution system analysis. From the software development perspective, it has been known that the inability to reuse code throughout the program and the difficulty in error checking using the procedure programming approach usually result in the software system being very difficult to maintain and extend.

From the power system simulation application perspective, there are three main challenges: 1) the simulation models and algorithms are difficult to extend; 2) the new computing hardware (e.g., multi-core CPU and GPU) and/or platforms (e.g., cloud computing) cannot be (fully) leveraged; 3) the simulation engines are not adequate to meet the analysis and simulation demand for the development of integrated power grid as well as integrated energy system. To address these challenges, a new generation power system simulation engine architecture and design is required. The InterPSS (Internet-technology Based Power System Simulator) project [1],[2] was created, aiming to provide an open simulation engine and the associated software development platform to the power engineering community where researchers and developers can easily extend the simulation engine or the platform to develop domain-specific or cross-domain power system simulation applications.

### B. Literature Review

Since there is limited information available regarding the internal architecture and development approach of the commercial (proprietary) simulation tools, the review is focused on the open-source simulation tools. For Matlab-based packages, there are PST [3], PSAT [4], and MatPower [5]. PYPOWER[6] and DOME[7] are developed based on Python. Open-source distribution system simulation tools include OpenDSS [8], developed in Delphi, and GridLAB-D [9], developed in C++. These tools focus on specific power system domain modeling and analysis functionality implementation. The tools are mainly developed as a "complete solution" package, with little focus on providing a platform or capabilities for extension and integration with other applications. It was noticed that the abstraction capability, modularity, and extensibility of object-oriented programming (OOP) was recognized by the developers of PSAT, MatPower, DOME, and GridLAB-D. Furthermore OOP concepts have been adopted in their development at various levels. However, no systematic and unified OOP approach is found, and the model (object) driven development approach not used in these open-source projects. Except for DOME having an independent core simulation module, other simulation tools are still heavily influenced by

This work was partially supported by the State Grid of China under the "Thousand Talents Plan" special research grant (5206001600A3).

the traditional software development approach. As a result, the extensibility and flexibility for integrating with other tools /platforms are inevitably limited.

*C. Contributions*

In summary, the following are the main contributions of the paper:

*1) Object Model*: InterPSS object model is fully object-oriented to support the model-driven development. In other words, the object model is placed at the center of InterPSS simulation engine. Everything else interacts with the central object model. The object model has been evolving over the last 20+ years [10] in a systematic and unified way. It has been extended recently to the distributed computing environment supporting in-memory computing.

*2) Model-Driven Development*: The model-driven development approach has been applied to the InterPSS simulation engine development. Particularly, Eclipse Modeling Framework (EMF) [11] based automated code generation technique has been successfully used in the InterPSS model development process. Through the use of automated software development technique, the software systems could be developed and evolved in a much more economic and timely manner, thereby significantly increasing both quality and productivity.

*3) Open Simulation Engine*: InterPSS simulation engine has a rather simple and loosely coupled open architecture. InterPSS object model provides APIs to interact with other parts of the simulation engine. Adapters are developed for data input to and result output from the object model. Simulation algorithms are decoupled from the object model, so that different algorithm implementations can be plugged-in. With this open architecture design, the engine can be easily extended, integrated with other software applications and adapted to new IT technologies.

*4) Engine Extension and Integration*: While traditional power system simulation software is designed to be a total solution package with the closed architecture, InterPSS simulation engine is designed to be an open generic power system simulation engine for extension and integration, with minimum GUI features. Several advanced applications, such as a hybrid simulation and an integrated transmission/ distribution system co-simulation, have been developed by extending and integrating with InterPSS simulation engine.

*D. Paper Organization*

Power system modeling requirements and associated modeling concepts are first discussed in Section-II. InterPSS object model is then introduced in Section-III, including the discussion of its inheritance relationship and extension mechanism. Model-driven software development approach is discussed in Section-IV, together with the Eclipse EMF technique used in the InterPSS model development process. InterPSS simulation engine is presented in Section-V, including the discussion of its system architecture and some of its salient features. Three InterPSS simulation engine based applications are discussed in Section-VI.

## II. POWER SYSTEM MODELING

Power system simulation software, except for those targeted for university teaching or academic research, is very complex and usually takes years to develop. When starting a large-scale software development project, the first important question is how to model the problem which the software is attempted to solve. This section discusses power system modeling requirement and associated modeling concepts from the power system simulation perspective.

*A. Modeling Requirement Analysis*

Traditionally power grid is modeled as a synchronous or "stiff" positive sequence network model in power system simulation. Asynchronous devices and un-symmetric events are modeled as an equivalent impedance or equivalent bus inject current to the positive sequence network model. With the increasing installation of power electronic devices and penetration of distributed generation and power electronic based loads, portions of a power grid might become inertia-less and much more dynamic. The positive sequence network model has inherent limitations in representing a large number of power electronic devices and more dynamic distribution systems in detail [16].

To model modern power grid with more dynamic and less-inertia behavior, the traditional positive sequence network model needs to be extended. The modeling issue can be viewed from two abstraction perspectives: 1) network modeling; 2) application modeling, as shown in Fig.1.

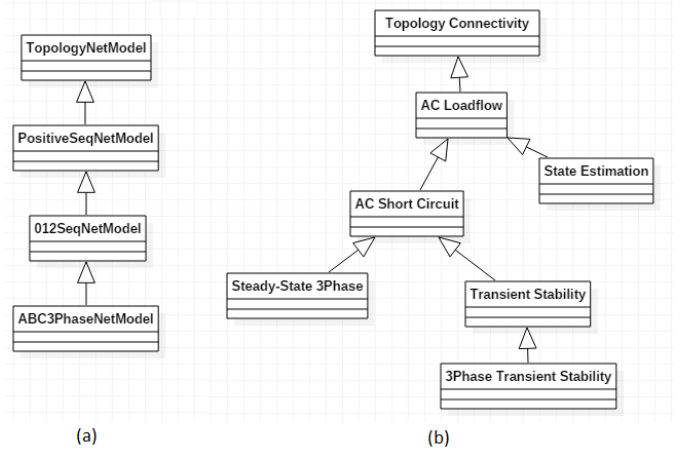

Fig. 1. Two Power System Modeling Perspectives, a) network modeling perspective; b) application modeling perspective, where the arrow symbol represents the object inheritance relationship.

From network modeling perspective, power grid model could be abstracted into 4 abstraction layers, as shown in Fig.1(a). In the figure, *TopologyNetModel* stands for topology connectivity network model; *PositiveSeqNetModel* for positive sequence network model; *012SeqNetModel* for 012 sequence coordinate network model; *ABC3PhaseNetModel* for ABC 3-phase coordinate network model. The functionality of these four model abstraction layers has the inheritance relationship for the reuse as shown in Fig.1.

The purpose of power grid modeling is to build power system simulation software, usually presented in the form of application programs for the user. From application modeling perspective, power grid model could be abstracted into multiple abstraction layers, as shown in Fig.1(b). In the figure, *Topological Connectivity* stands for power network topology analysis; *AC Loadflow* for AC loadflow calculation; *AC Short Circuit* for AC short circuit analysis; *State Estimation* for state estimation calculation; *Steady-State 3Phase* for ABC 3-phase steady-state analysis; *Transient Stability* for dynamic (transient) stability simulation; *3Phase Transient Stability* for ABC 3-phase dynamic (transient) stability simulation.

We believe that the abstraction concepts, as shown in Fig.1, have captured the modeling requirements and, therefore, provide a good foundation for modeling modern power grid with more dynamic and less-inertia behavior. Additionally, the decoupling of the abstract/functional modeling concepts and the inheritance/reuse relationship could provide guidance for developing more maintainable, reusable, flexible and extendible power system simulation software.

### B. Conceptual and Implementation Model

A model, in the most general sense, is anything used in any way to represent anything else. There are different types of models for different purposes, such as physical model, conceptual model and concrete implementation model [12]. A conceptual model may only be drawn using some artifacts, which is commonly used to help us to get to know and understand the subject matter it represents. A conceptual model could be implemented in multiple ways. On the other hand, a concrete implementation model, in general, is precise in terms of computer implementation, which can be translated to the corresponding software representation directly and possible automatically. Some models are both conceptual and concrete. For example, data models defined using XML schema are both conceptual and concrete models.

### C. Data Processing Pattern

A power system simulation application, in general, is intended to process some input information by applying some algorithms to produce some output results. Broadly speaking, there are two application data processing patterns: 1) algorithm focused pattern; 2) model focused pattern, as shown in Fig.2. In the algorithm focused pattern (Fig.2(a)), the input information is first cached in a set of temporary array variables. Then the input is used to call some algorithm function(s) to produce the result. The result is usually cached in another set of temporary array variables again before being formatted to produce the output. In the model focused pattern (Fig.2(b)), objects are first created in the computer memory. The input information is stored in the object model, some algorithm is then applied to the object to produce the result, which is again stored in the same object before being formatted to produce the output.

When using the procedure programming approach, the algorithm focused pattern is employed, while the model focused pattern is employed when using the object-oriented programming approach. InterPSS simulation engine is fully object-oriented and therefore is based on the model focused data processing pattern.

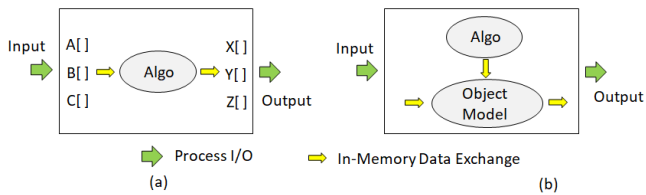

Fig. 2. Comparison of the algorithm focused pattern (a) and the model focused pattern (b).

### D. Node/Breaker Model and Bus/Branch Model

Power system simulation model could be, in general, categorized into two types: 1) node/breaker physical device model; 2) bus/branch logical simulation model. IEC common information model (CIM) [13] is a node/breaker data model, designed to model/describe physical devices and their connectivity. Information, such as physical equipment connectivity, equipment physical size, equipment geographical location (attitude and longitude), as well as equipment electrical parameters (R, X, Rating), are included in the model. As we know, almost all modern power system simulation algorithms are formulated on the node admittance Y-matrix. Bus, Branch and its connectivity information are the foundation of a Bus/Branch model. The CIM model does not have direct representation of the buses, branches, and its connectivity information, and therefore, could not be used to efficiently construct a bus/branch model.

The IEEE working group first attempted to specify a bus/branch model by defining the IEEE Common Data Format (IEEE CDF) for exchanging Loadflow study data [14] using flat files. Currently, power system software vendors use their own internal data format, such as PSS/E, BPA or PowerWorld format, to exchange power system simulation information. The Open Data Model for Exchange Power System Simulation Data (ODM) [15] is an effort to create a common data format for exchange bus/branch model information. This effort has been leveraged by InterPSS as the default approach for importing data from different formats [12].

InterPSS simulation engine is intended to meet the new modeling requirements outlined in this section. It is based on the model focused data processing pattern. InterPSS object model, discussed in the next section, is a concrete bus/branch implementation model, which has a precise definition as for how it is represented in the computer memory and provides a programming API to interact with the outside world.

### III. INTERPSS OBJECT MODEL

In object-oriented programming, the main parts in a program are objects. The object-oriented approach lets you create objects that model real-world objects. Programs are made up of parts, which can be coded and tested separately, and then assembled to form a complete application. At the center of InterPSS simulation engine is its object model, which has been evolving over the last 20+ years [10]. This section discussed InterPSS object model and how it was designed, and might be extended to support multi-domain/scale power system modeling and simulation.

### A. Model Extension Convention

InterPSS object model follows a unified extension conversion as shown in Fig.3. Classes with pre-fix base classes (*BaseClass, BaseSubClass*) are abstract, which are used to defined concrete application classes (*AppClass*). Some of the fields in the base-class are with generic types. The Field type template (*TField*) is declared for the generic types. When defining concrete application classes (*AppClass*), the field type template is substituted by some concrete field class definition (*FieldClass*).

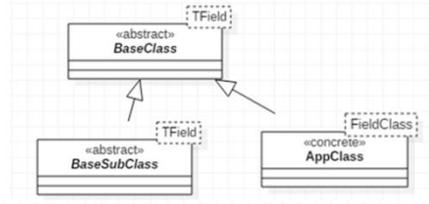

Fig. 3. InterPSS Class Extension Mechanism

### B. InterPSS Object Model

InterPSS object model was first implemented in C++ in 1996 [10] and then ported to Java in 2005 [2]. Currently, the main network concept is modeled using an object class structure as shown in Fig.4. In the figure, *AclfNetwork/AclfBus/AclfBranch* are for AC loadflow calculation; *AcscNetwork/AcscBus/AcscBranch* for AC short circuit analysis; *Acsc3PNetwork/Bus/Branch* for ABC 3-phase steady-state analysis; *DStabilityNetwork/DStabBus/DStabBranch* for dynamic (transient) stability simulation; *DStab3PNetwork/DStab3PBus/DStab3PBranch* for ABC 3-phase dynamic (transient) stability simulation. *AclfNetwork* is inherited from the *BaseAclfNetwork* class with the substitution of the *TBus/TBranch* template by *AclfBus/AclfBranch*, indicating that *AclfNetwork* contains *AclfBus/AclfBranch* objects to form an AC loadflow network model.

InterPSS object model supports modeling multiple networks of different types, for example, a mix of positive sequence, 012 sequence, ABC 3-phase network models in a simulation application model. As shown in Fig.4, a parent *Network* class might contain 0 to many child networks (*childNet*). The relationship is nested, which means a child network, in turn, can contain its grand-child networks.

In theory, there is no difference in the algorithm used in transmission system analysis and distribution system analysis. In practice, the input and output data might have different formats for transmission and distribution systems. InterPSS distribution system analysis network class *DistNetwork* hierarchy is shown in Fig.5. A *DistNetwork* class is a subclass of the *Network* class. It also contains an *AcscNetwork* object for distribution system loadflow analysis and short circuit calculation.

Over the last 20 years, since the InterPSS object was first defined in [10], we have seen significant technology changes in the modern power grid. It has been becoming much more dynamic and less synchronized with increasing installation of power electronic devices and the increasing penetration of distributed generation and loads. InterPSS object model has a loosely coupled open architecture, which allows the continuous and smooth evolution of the model to address the challenges due to the technology advances both in the IT world and the power engineering world.

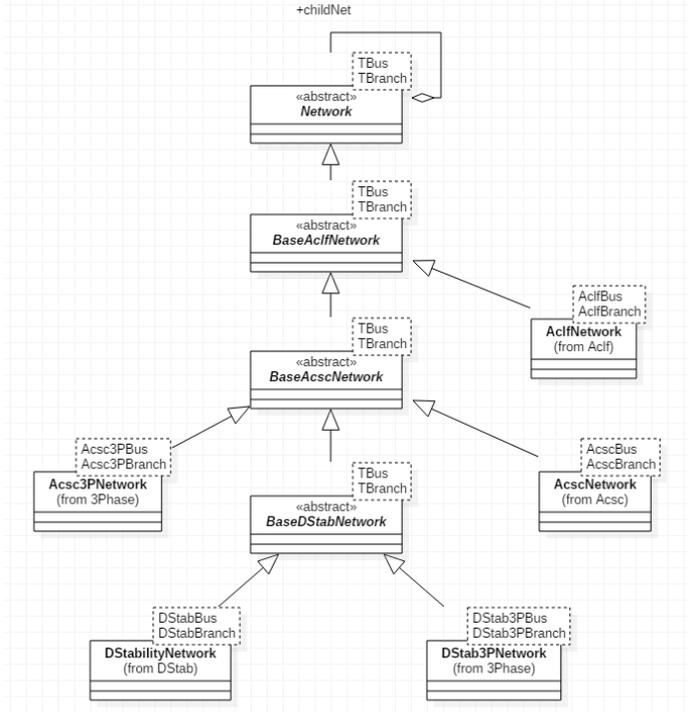

Fig. 4. InterPSS Network Class Hierarchy, where diamond symbol represents object composition relationship.

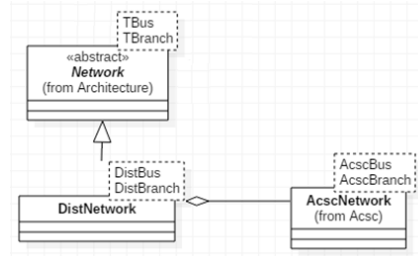

Fig. 5. InterPSS Distribution System Analysis Network Class Hierarchy

## IV. MODEL-DRIVEN DEVELOPMENT

Model-driven architecture (MDA) [17] is a design approach for the software development. In this approach, software specifications are expressed as conceptual models, which are then "translated" into one or more concrete implementation models. The translation is often performed by automated tools.

### A. Eclipse Modeling Framework

Eclipse Modeling Framework (EMF) [11] is a modeling framework and code generation facility for building tools and applications based on structured data models. From a model specification, EMF provides tools and runtime support to produce a set of Java classes for concrete implementation of the model. EMF is a subset of the UML specification. Models defined using UML are conceptual models, while models defined using EMF are concrete implementation models. IEC CIM is specified in UML and therefore is a conceptual model. As a consequence, different vendors have different concrete implementations of CIM. InterPSS object model, defined using EMF, is a concrete implementation model.

Snapshots of EMF representation of InterPSS object model are shown in Fig.6, where Fig.6(a) is a package view of the *core* package, and Fig.6(b) a class view of the *AclfNetwork* class. The inheritance relationship *between AclfNetwork* and *BaseAclfNetwork* is also shown in the figure. After defining the EMF model, InterPSS object model code is automatically generated from the EMF definition.

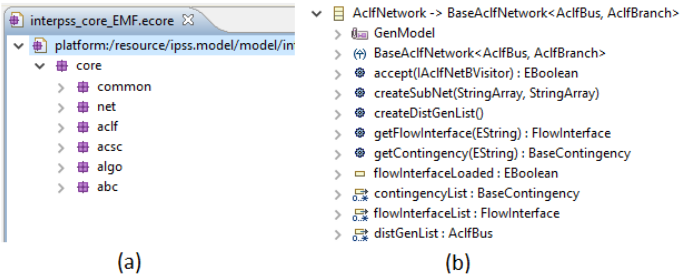

Fig. 6. Sample screen shots of the EMF representation of InterPSS object model.

### B. InterPSS Development Process

MDA based software development approach is sometimes called model-driven development approach. In the approach, the model is placed at the center of software development process and the rest evolve around the model. The following is a summary of the model-driven development approach used by InterPSS:

1) InterPSS object model is first defined in Eclipse EMF as a conceptual model. Since the conceptual model is represented by a set of UML-classes, it is intuitive and quite easy to maintain.

2) Using Eclipse EMF code generation tool, the conceptual model is translated to a set of Java classes, representing the concrete InterPSS implementation model, which is used to represent the object model in the computer memory for the simulation purposes.

3) The object model provides an API [18]. Other InterPSS simulation engine components interact with the object model through the API.

Today power system simulation software systems are significantly large and complex. Through the use of the automated software development approach, the software systems could be developed and evolve in a much more economic and timely manner, thereby significantly increasing both quality and productivity. Eclipse EMF based automated code generation approach has been successfully used in the InterPSS model development process. It is estimated that currently more than 50% of InterPSS object model source code is auto-generated by the computer.

## V. INTERPSS SIMULATION ENGINE

InterPSS power system simulation engine is based on InterPSS object model.

### A. Engine Architecture

InterPSS simulation engine has a rather simple and loosely coupled architecture, as shown in Fig.7. Please note: the information/data flow chat is commonly used to explain the main idea as for how a simulation algorithm/function is implemented, in general, with a start-point and an end-point. The relationship depicted in Fig.7 is at a much higher software architecture level, explaining how the main simulation engine components (objects) are interacting with each other. InterPSS object model provides APIs to interact with other parts of the simulation engine. Adapters are developed for data input to and result output from the object model. For example, an ODM adapter was developed to input simulation data stored in the DOM format into the object model [19]. Simulation algorithms are decoupled from the object model so that different simulation algorithm implementations can be plugged-in through the software configuration.

The simulation engine supports the implementation of two types of algorithms as shown in Fig.7:

1) Mutable Algorithm *(Algo in Fig.7)*: When this type of algorithms are applied to the object model, they change (mute) the state of the object model. For example, during loadflow calculation, the algorithm changes the bus voltage magnitude/angle, attempting to minimize the bus power mismatch [10];

2) Immutable Algorithm *(Algo(P) in Fig.7 )*: When this type of algorithms are applied, the object model is immutable or in the ready-only mode, that is the algorithm only reads data from the object model and never writes back. This type of algorithms are multi-thread safe and can apply to the object model in parallel.

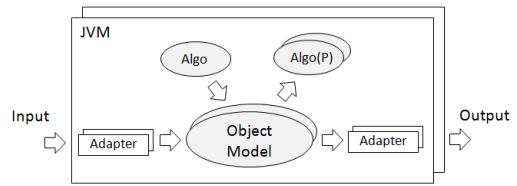

Fig. 7. InterPSS simulation engine high-level architecture view.

### B. Engine Runtime Environment

InterPSS simulation engine is written in the Java programming language, and therefore can be run in any place where Java Virtual Machine (JVM) runtime environment is available. A desktop PC version of InterPSS application was first developed and made available to download freely in 2006 [1]. Then a Cloud edition of InterPSS application was released in 2009 [12],[20], which is running in the Google Cloud and can be accessible anywhere 24x7. InterPSS simulation engine was ported to Gridgain [29] to support grid computing [12],[21] and Hazelcast [30] to support in-memory computing [22] in the distributed computing environment.

*C. Parallel Computing*

InterPSS simulation engine supports parallel computing at three levels:

1) *Multi-JVM level*: InterPSS simulation engine can be distributed to multiple JVMs and runs in parallel. If these JVMs are managed by a data grid software, such as Gridgain [29] or Hazelcast [30], the distribution of InterPSS object model to the computer nodes in a cluster is automatic.

2) *Multi-case level*: InterPSS object model is multi-thread safe. Therefore, multiple objects, each representing an independent simulation case, could reside in one JVM simultaneously. These simulation objects could be processed in parallel.

3) *Algorithm level*: Certain types of simulation algorithms could be designed and implemented such that the object model is immutable or in read-only mode during the execution of the algorithm. The algorithm implemented in such a way is multi-thread safe and can apply to the object model in parallel.

In summary, InterPSS simulation engine supports parallel implementation at three levels: application/program (computer process) level, object model level, and simulation algorithm level. The algorithm level parallelism is most efficient among the three parallel computing approaches. For example, in the N-1 contingency analysis (CA), after factorization of the pre-contingency network B-matrix, contingency analysis could be designed to run in parallel. It should be pointed out that not all power system simulation algorithms can be designed to achieve the algorithm level parallelism. Using a real-world large-scale online power network model (40K+ buses) as a test case, the total computation time for a complete N-1 CA of the power network could be achieved in less than 2 seconds with the parallel approach, and with an acceleration ratio of approximate 29 times as compared with the sequential (no parallelism) approach [22]. The testing hardware used was a high-performance server machine with 4 CPU and total 40 cores.

*D. In-Memory Computing*

In-memory computing (IMC) means using the data grid that allows one to store data in computer RAM, across a cluster of computer nodes, and process it in parallel. IMC software, in general, is designed to store data in a distributed fashion, where the entire dataset is divided and stored in individual computer nodes' RAM, each storing only a portion of the overall dataset. Once data is partitioned, highly efficient parallel distributed processing becomes possible. An IMC technology based power grid analysis approach has been developed [21],[32], which is implemented on top of InterPSS object model.

InterPSS simulation engine is based on InterPSS object model, and has a rather simple and loosely coupled open architecture. InterPSS object model provides APIs to interact with other parts of the simulation engine. Simulation algorithms are decoupled from the object model, so that different algorithm implementations can be plugged-in. With this architecture design, the engine can be easily adapted to new computing hardware or platform, extended and integrated with other applications.

VI. ENGINE EXTENSION AND INTEGRATION

This section presents some examples where InterPSS simulation engine has been extended or integrated with other tools to realize advanced simulation features and solve challenging simulation problems.

*A. Integrated Transmission and Distribution Co-simulation*

The interactions between distribution and transmission systems have increased significantly in recent years. However, in traditional power system simulation tools, transmission and distribution systems are separately modeled and analyzed. Hence, it is difficult to analyze the impacts of distribution systems on transmission systems and their interactions in detail. To fill the gap, integrated transmission and distribution (T&D) power flow and dynamic co-simulation algorithms have been developed by extending InterPSS simulation engine [23].

In the T&D co-simulation implementation, each transmission or distribution system is represented by one InterPSS network object model (see Fig.4). Thanks to the model-algorithm decoupled architecture, each network can be solved individually and their solutions can be flexibly exchanged and coordinated. The major work in this extension exercise focus on developing algorithms to coordinate the individual network solution.

At each iteration of during the loadflow calculation, the transmission system provides the three-phase voltages at the boundary buses, denoted by $V_{B,i}^{abc}$, to the corresponding distribution systems to update their boundary bus voltages. The distribution systems send their three-sequence equivalents (equivalent load for positive sequence, equivalent current injections for both negative- and zero-sequence) to the transmission system. The data exchange process is illustrated in Fig. 8.

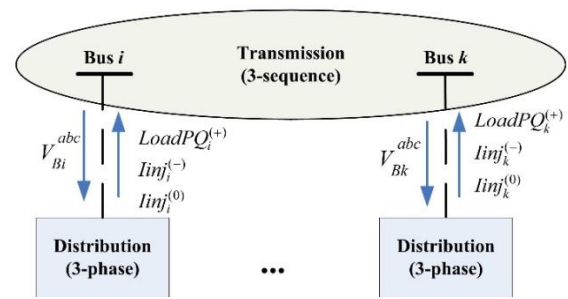

Fig. 8. Data exchange for integrated T&D power flow solution

During the dynamic simulation, the Multi-area Thévenin Equivalent (MATE) approach [23] is employed to solve the network solution step, which is illustrated in Fig. 9.

Transmission and distribution system are solved independently at each integration step.

The developed integrated T&D co-simulation program has been tested on a large-scale integrated T&D system which consists of a modified IEEE 300-bus system [24], and 43 distribution systems (for the 43 load buses with loads greater than 50 MW each in the area 1 of the transmission system) that are built based on the IEEE 13-bus feeder [25]. The T&D system has combined 22920 buses in total with 1740 feeders.

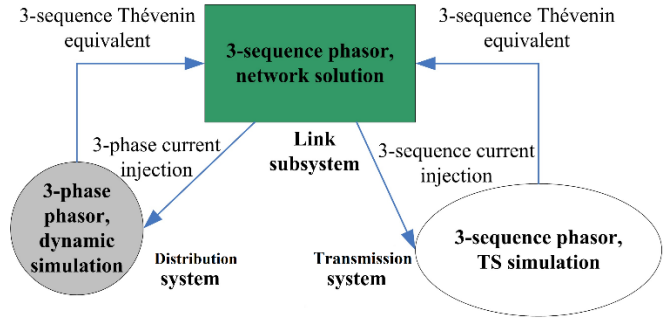

Fig. 9. The MATE based network solution for integrated T&D dynamic co-simulation

The following is a sample dynamic simulation. At 1.0 sec, a three-phase to ground fault is applied to the low voltage side of the step-down transformer connected to Bus 2 of the transmission system, and the fault is cleared after 0.07 sec. The dynamic simulation results for the T&D system are shown in Figs. 10 and 11.

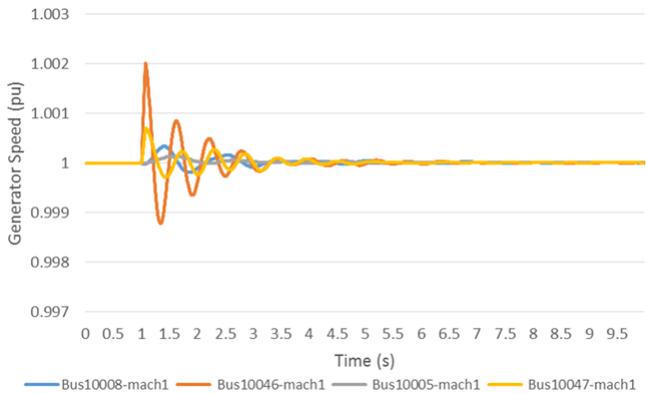

Fig. 10 The speeds of four generators in area 1 of the transmission system

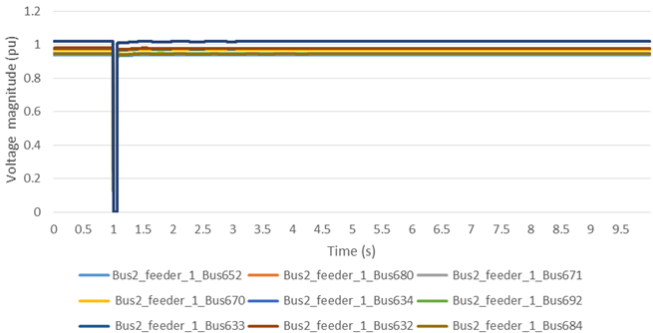

Fig. 11. The phase *A* line-to-neutral voltages of buses of a feeder served by Bus 2 of the transmission system.

### B. Integration with a Market Simulator

The Market Simulator is a software tool, developed by an Independent System Operator (ISO) in the US, intended for building quick prototypes for studying/analyzing different electricity market models and optimization methodologies /algorithms. To achieve the goal of building quick model /algorithm prototypes, the simulator optimization engine is implemented using GAMS [26]. The simulator is responsible for the optimization part of the solution algorithms, but also requires some of the power network analysis functionality (topology processing, contingency/sensitivity analysis). InterPSS simulation engine was integrated with GAMS portion of the simulator to provide the network analysis functionality. InterPSS and the simulator integration architecture is shown in Fig.12, where the integration is through in-memory API calls, inside the JVM.

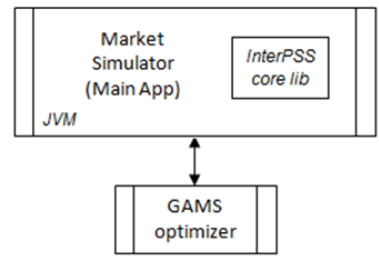

Fig. 12. InterPSS and the Market Simulator integration architecture.

### C. Hybrid Simulation

InterPSS simulation engine was integrated with PSCAD /EMTDC through TCP/IP socket communication to create an Electromagnetic Transient (EMT) and Transient Stability (TS) hybrid simulation platform, as shown in Fig. 13. For the integration, three new modules including the socket server, the network equivalent helper, and the hybrid simulation manager were developed [27]. Three-phase modeling in ABC coordinate and three-sequence modeling in 012 coordinate TS simulation features of InterPSS simulation engine facilitated the integration with the EMT simulators, especially under the unbalanced system condition. The same integration approach could also be used to interface with other mainstream EMT simulators.

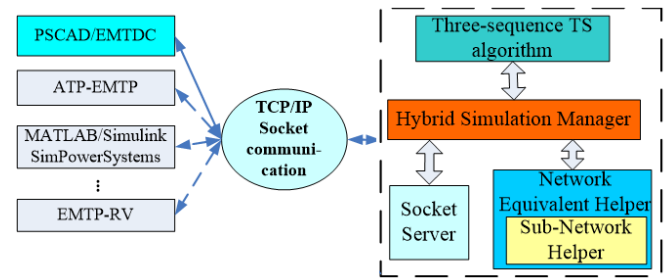

Fig. 13 Architecture of the developed hybrid simulation

This hybrid simulation platform has been applied to detailed VSC-HVDC simulation on a modified IEEE 39-bus test system [28] and the fault-induced delayed voltage recovery (FIDVR) analysis of a large-scale power system with 15750 buses, of which 238 were modeled in detail using an EMT simulator [27].

## VII. CONCLUSIONS

Traditionally, power system simulation tool development focuses on various analysis algorithm/functionality development, while little attention is paid to the software system architecture, extensibility and the integration capability. The recent development of smart grids and integrated energy systems calls for the trans-disciplinary and multi-domain simulation, where one power system domain simulation need to be integrated with other applications/tools, possibly of the other domains. In light of this, the paper introduced the design, development, and application of InterPSS simulation engine, which has an open and loosely coupled software system architecture. This architecture enables components developed by others to be easily plugged into InterPSS to augment its functionality, and equally important, allows the engine to be integrated with other software systems through extension and integration.

The model-driven development approach, especially the Eclipse Modeling Framework (EMF) based automated code generation technique, has been successfully used in the InterPSS object model development process. The InterPSS project over the years has been developed and maintained by a group of volunteers. The use of the automated software development approach is one of the major contributing factors to the significantly increased software quality and productivity.

Although the original object-oriented power grid modeling idea was proposed some 20+ year ago [4], due to the loosely coupled open architecture design, new IT technologies, such as distributed data grid and in-memory computing, have been integrated into InterPSS simulation engine, and the simulation engine has been extended to run in the distributed computing environment and the Cloud. InterPSS simulation engine recently has been integrated with Google's TensorFlow machine learning (ML) engine to create an open platform for exploring the application of ML to power system analysis [31]. Several advanced applications, such as an integrated transmission/distribution system co-simulation, a market simulator, and a hybrid simulation platform have been developed by others based on extending InterPSS simulation engine and/or integrating it with other applications/tools.